\documentclass{article}
\usepackage{fleqn,espcrc2}

\usepackage{graphicx}
\usepackage{url}

\title{Numerical Modelling of Optical Trapping}

\author{T. A. Nieminen, H. Rubinsztein-Dunlop,
	N. R. Heckenberg and A. I. Bishop
	\address{Centre for Laser Science, Department of Physics,\\
		The University of Queensland, Brisbane QLD 4072, Australia}
}

\begin{document}

\begin{abstract}
Optical trapping is a widely used technique, with many important
applications in biology and metrology. Complete modelling of trapping
requires calculation of optical forces, primarily a scattering problem,
and non-optical forces. The T-matrix method is used to calculate forces
acting on spheroidal and cylindrical particles.

\vspace{-6.5cm}
{\small\noindent
\textbf{Preprint of:}\\
T. A. Nieminen, H. Rubinsztein-Dunlop,
N. R. Heckenberg and A. I. Bishop\\
``Numerical Modelling of Optical Trapping''\\
\textit{Computer Physics Communications} \textbf{142}, 468--471 (2001)
}
\vspace{5.2cm}


\vspace{1pc}
\end{abstract}

\maketitle

\section{Introduction}

Optical trapping provide three-dimensional confinement and
manipulation of microscopic particles by a focussed laser beam.
Optical trapping is a powerful and widespread technique, with the
single-beam gradient trap (also known as optical tweezers) in use for a
large number of biological and other applications.

\begin{figure}[htb]
\centerline{\includegraphics[width=7.5cm]{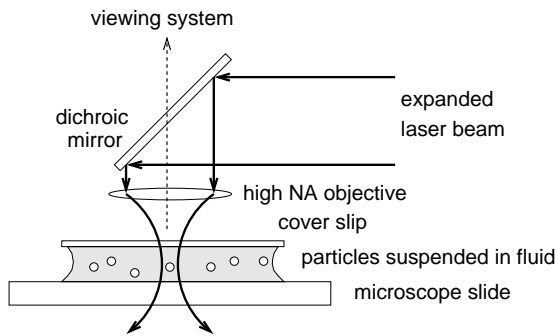}}
\caption{Schematic diagram of a typical optical tweezers setup}
\label{tweezersfig}
\end{figure}

The trapping beam applies optical forces (usually divided into a
gradient force, acting towards areas of higher irradiance, and
scattering forces, including absorption and reflection forces)
to the particle.

The optical forces and torques result from the transfer
of momentum and angular momentum from the trapping beam to the particle.
Various approximate methods such as geometric optics or
Rayleigh approximations are often used for the calculation of the
optical forces. Such approximate methods are not necessary, since
electromagnetic scattering theory can be used for the calculation of
forces, avoiding the limited ranges of applicability of the approximate
methods.

Other forces will also affect the motion of the particle. The most
important of these forces, gravity, bouyancy, and viscous drag as the
particle moves through the surrounding fluid, are readily taken into
account.

\section{Trapping as a scattering problem}

The optical forces and torques applied to the particle result from the
transfer of momentum and angular momentum from the trapping beam to the
particle. The total momentum transfer can be found by solving the
electromagnetic scattering problem. A variety of numerical methods can
be used --
finite element method, FDTD, discrete dipole
approximation~\cite{draine1994}, the T-matrix
method~\cite{mishchenko1991,tsang1985}, etc.

A number of these have been used for optical force calculations,
including forces in optical traps~\cite{white2000,kimura1998}. One
method, however, stands out as ideal for trapping force calculations --
the T-matrix method. The T-matrix method can be considered an
extension of Mie theory to arbitrarily shaped particles with arbitrary
illumination. The main advantage of the T-matrix method is that trapping
calculations usually involve repeated calculation of the scattering for
the same particle under differing illumination. In this case, the
T-matrix need only be calculated once, since it is independent of the
fields, whereas methods such as FEM, FDTD and DDA will require the
entire calculation to be repeated.

In the T-matrix method, the incident trapping field illuminating the
particle is expressed as a sum of regular vector spherical wave
functions (VSWFs):
\begin{eqnarray}
\mathbf{E}_{\mathrm{inc}}(\mathbf{r})  =
\sum_{n=1}^\infty \sum_{m=-n}^n
[ a_{mn} \mathbf{RgM}_{mn}(k\mathbf{r}) +
& & \nonumber \\
b_{mn} \mathbf{RgN}_{mn}(k\mathbf{r}) ] & &
\label{incidentfield}
\end{eqnarray}
where
\begin{eqnarray}
\mathbf{RgM}_{mn}(k\mathbf{r}) & = &
	(-1)^m d_n  \exp(im\phi) \times
\nonumber \\ & &
j_n(kr) \mathbf{C}_{mn}(\theta),
\label{RgM}
\end{eqnarray}
\begin{eqnarray}
\mathbf{RgN}_{mn}(k\mathbf{r}) =
	(-1)^m d_n \exp(im\phi) \times
& & \nonumber \\
	\left\{ \frac{n(n+1)}{kr} j_n(kr)
	\mathbf{P}_{mn}(\theta) + \right.
& & \nonumber \\
	 \left. \left[j_{n-1}(kr) - \frac{n}{kr}j_n(kr) \right]
	  \mathbf{B}_{mn}(\theta) \right\}, & &
\label{RgN}
\end{eqnarray}
\begin{equation}
\mathbf{B}_{mn}(\theta) =
	\hat\theta \frac{d}{d\theta} d_{0m}^n(\theta)
	+ \hat\phi \frac{im}{\sin\theta} d_{0m}^n(\theta),
\end{equation}
\begin{equation}
\mathbf{C}_{mn}(\theta) =
	\hat\theta \frac{im}{\sin\theta} d_{0m}^n(\theta)
	- \hat\phi \frac{d}{d\theta} d_{0m}^n(\theta),
\end{equation}
\begin{equation}
\mathbf{P}_{mn}(\theta) = \hat{r} d_{0m}^n(\theta),
\end{equation}
\begin{equation}
d_n = \left( \frac{2n+1}{4\pi n (n+1)} \right)^{\frac{1}{2}},
\end{equation}
$j_n(kr)$ are spherical Bessel functions, and $d_{0m}^n(\theta)$ are
Wigner $d$ functions.

Similarly, the scattered fields are expressed as a VSWF expansion.
In this case, since the far field must be an outgoing radiation field,
\begin{eqnarray}
\mathbf{E}_{\mathrm{scat}}(\mathbf{r}) = \sum_{n=1}^\infty
	\sum_{m=-n}^n 	[ p_{mn} \mathbf{M}_{mn}(k\mathbf{r})
& & \nonumber \\
	+ q_{mn} \mathbf{N}_{mn}(k\mathbf{r}) ] & &
\end{eqnarray}
where $\mathbf{M}_{mn}(k\mathbf{r})$
and $\mathbf{N}_{mn}(k\mathbf{r})$ are the same as
$\mathbf{RgM}_{mn}(k\mathbf{r})$ and
$\mathbf{RgN}_{mn}(k\mathbf{r})$, with the spherical Bessel
functions replaced by spherical Hankel functions of the first
kind, $h_n^{(1)}(kr)$.

From the linearity of the Maxwell equations, there is a linear
relationship between the incident and scattered fields:
\begin{equation}
p_{mn} = \sum_{m'n'} T_{mnm'n'}^{(11)} a_{m'n'}
	+ T_{mnm'n'}^{(12)} b_{m'n'}
\end{equation}
\begin{equation}
q_{mn} = \sum_{m'n'} T_{mnm'n'}^{(21)} a_{m'n'}
	+ T_{mnm'n'}^{(22)} b_{m'n'}
\end{equation}
The T-matrix can be calculated using the extended boundary condition
method (EBCM)~\cite{mishchenko1991,tsang1985}. For spherical particles,
the T-matrix becomes diagonal, and the non-zero elements are the usual
Mie coefficients. For rotationally symmetric particles, the T-matrix is
diagonal with respect to $n$. Computer codes to calculate T-matrices for
such rotationally symmetric particles are
available~\cite{mishchenkocode}.

\section{Representation of the trapping beam}

The use of the T-matrix method for scattering calculations requires that
the trapping beam be represented in terms of vector spherical
wave functions, that is, the coefficients $a_{mn}$ and $b_{mn}$ in
equation~(\ref{incidentfield}) need to be found. The regular VSWFs
$\mathbf{RgM}_{mn}$ and $\mathbf{RgN}_{mn}$ provide a complete set of
modes or partial waves, each individually satisfying the Maxwell
equations, which can be used to represent any incident electromagnetic
wave. For the simple case of a plane wave,
$\mathbf{E}(\mathbf{r}) = \mathbf{E}_0 \exp(i\mathbf{k}\cdot\mathbf{r})$,
with $\mathbf{k}$ in the direction
$(\theta,\phi)$, the expansion coefficients
are~\cite{mishchenko1991,tsang1985} \begin{equation}
a_{mn} = 4 \pi (-1)^m i^n d_n
	\mathbf{C}_{mn}^\star \cdot \mathbf{E}_0 \exp(-im\phi)
\label{planea}
\end{equation}
\begin{equation}
b_{mn} = 4 \pi (-1)^m i^{n-1} d_n
	\mathbf{B}_{mn}^\star \cdot \mathbf{E}_0 \exp(-im\phi).
\label{planeb}
\end{equation}
Note that the amplitude vector $\mathbf{E}_0$ contains the information
regarding the polarisation and phase of the wave, and can be complex.

In an optical trap, the incident field is usually a strongly focussed
Gaussian or other beam. In principle, such a beam can either be
decomposed directly into a VSWF representation, or into a plane wave
spectrum, from which the VSWF expansion coefficients can be found using
equations (\ref{planea}) and (\ref{planeb}). In practice, this is
problematic, since the usual descriptions of beams do not actually
satisfy the Maxwell equations.

For the case of Gaussian beams,
either plane wave expansion~\cite{doicu1997} or direct VSWF
expansion~\cite{lock1994,gouesbet1994} can be
used, although neither will give a beam identical to a traditional
Gaussian beam.

\section{Optical forces}

Using the T-matrix method, with the T-matrix calculated by the
publically available code by Mishchenko~\cite{mishchenkocode}, and the
beam shape coefficients in the localised approximation by
Gouesbet~\cite{lock1994,gouesbet1994} used to describe the beam, we
calculated the variation of the axial force acting on particles of
varying shape as a function of their position along the beam axis.

\begin{figure}[!htbp]
\centerline{\includegraphics[width=7.5cm]{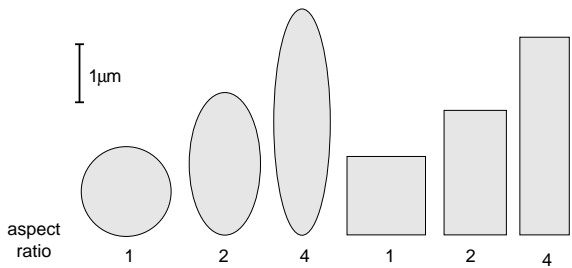}}
\caption{Differently shaped spheroidal and cylindrical particles with
aspect ratios of 1, 2, and 4.}
\label{particles}
\end{figure}

The particles are polystyrene ($n = 1.59$) prolate spheroids and
cylinders, of varying aspect ratio as indicated (see
figure~\ref{particles}). The particles are of equal volume, with a
volume equal to that of a sphere of radius 0.75$\mu$m, and are
trapped in water by a Gaussian beam of waist width 0.8$\mu$m and free
space wavelength 1064nm.

\begin{figure}[!htbp]
\centerline{\includegraphics[width=7.5cm]{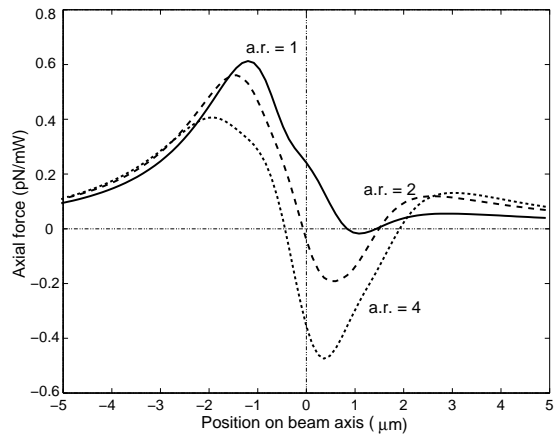}}
\caption{Axial force acting on spheroids of aspect ratios 1, 2, and 4.}
\label{spheroids}
\end{figure}

\begin{figure}[!htbp]
\centerline{\includegraphics[width=7.5cm]{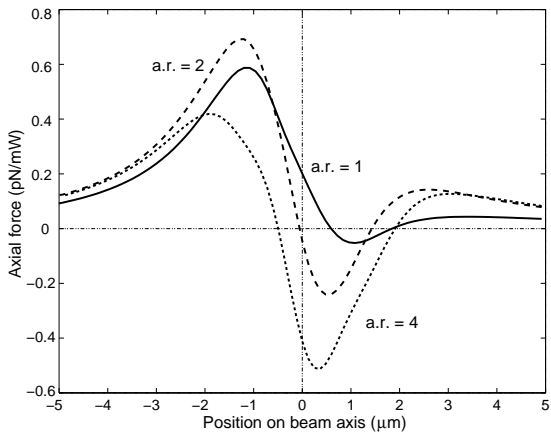}}
\caption{Axial force acting on cylinders of aspect ratios 1, 2, and 4.}
\label{cylinders}
\end{figure}

The axial forces acting on the spheroids and cylinders are shown
in figures~\ref{spheroids} and \ref{cylinders}. A negative position on
the beam axis indicates a position before the focal plane is reached, a
positive position is after the focus. A positive force acts to push the
particle in the direction of propagation of the beam, a negative force
will act to axially trap the particle. If only optical forces are
acting, the particle will come to rest at the zero optical force
position where force curve crosses the zero force line with a negative
gradient.

The trapping forces acting on the spheroid and cylinder with aspect
ratio 1 is very small. This is due to interference
due to reflections from the rear surface of the
particle~\cite{maianeto2000}.

As the particles become smaller, the force becomes less dependent on
the shape of the particle. Even for the particles considered here, it
can be seen that the fine details of the shape (i.e. spheroid vs
cylinder) only has a small effect on the force. As the particle becomes
more elongated, it obstructs less of the trapping beam, and radiation
pressure forces are reduced.

\section{Non-optical forces}

A number of non-optical forces will act on the particle. Buoyancy and
gravity are constant and are simply dealt with. Motion of the particle
in the surrounding fluid is completely dominated by viscous drag.

Since the time required for the particle to reach
its terminal velocity of $\approx 1 \mu\mathrm{ms}^{-1}$
is very short ($\tau \approx 10^{-7}\mathrm{s}$ is typical), we
use $\dot{\mathbf{r}} \propto \mathbf{F}$
instead of $\ddot{\mathbf{r}} \propto \mathbf{F}$,
since the particle will be moving at very close to the terminal velocity
at all times. If the fluid is in motion, we use the velocity of the
particle relative to the fluid. Since typical Reynolds numbers in
trapping are extremely low (Re~$\approx 10^{-5}$ is typical), perfect
laminar flow can be assumed -- for a spherical particle, Stoke's Law
will be an excellent approximation.

In general, the trapping beam will heat the surrounding medium,
possibly giving rise to convective flow. While free convection is, in
general, a difficult problem, the convection problem in trapping is
perhaps the simplest possible. Due to the very small distances involved,
the steady-state temperature distribution is reached in a very short
time~\cite{quantumchemistry}, and a steady-state temperature
distribution, \emph{independent of the convective flow}, can be assumed.
Similarly, any convective flow will reach steady-state very
quickly.

Further effects that can be included for a complete model include
Brownian motion, thermophoretic effects due to uneven heating of the
particle, the effects of nearby particles and surfaces, etc.
Accurate calculation of the optical forces can allow the evaluation of
the accuracy of the modelling of these effects.

\end{document}